\documentclass[a4paper,twocolumn,
english,aps,prb,floatfix,showpacs]{revtex4}
\usepackage[T1]{fontenc}
\usepackage[latin1]{inputenc}
\usepackage{amsmath}
\usepackage{babel}
\usepackage{graphics}
\usepackage{amssymb}

\begin{document}
 
\title{Multicritical point of Ising spin glasses on triangular and
honeycomb lattices
}

\author{S.L.A. \surname{de Queiroz}}

\email{sldq@if.ufrj.br}

\affiliation{Instituto de F\'\i sica, Universidade Federal do
Rio de Janeiro, Caixa Postal 68528, 21941-972
Rio de Janeiro RJ, Brazil}

\date{\today}

\begin{abstract}
The behavior of two-dimensional Ising spin glasses at the multicritical point on 
triangular and  honeycomb  lattices is investigated, with the help of finite-size scaling 
and conformal-invariance concepts. We use transfer-matrix methods on long strips to 
calculate domain-wall  energies,  uniform  susceptibilities, and spin-spin  correlation 
functions. Accurate estimates are provided for the location of the 
multicritical point on both lattices, which lend strong support to a conjecture recently 
advanced by Takeda, Sasamoto, and Nishimori. Correlation functions are shown to
obey rather strict conformal-invariance requirements, once suitable adaptations are made 
to account for geometric aspects of the transfer-matrix description  of triangular and  
honeycomb  lattices.
The universality class of critical behavior
upon crossing the ferro-paramagnetic phase boundary is probed, with
the following estimates for the associated critical indices: 
$\nu=1.49(2)$, $\gamma=2.71(4)$, $\eta_1= 0.183(3)$, distinctly different from the 
percolation values.  
\end{abstract}
\maketitle
\section{INTRODUCTION}
\label{intro}
In this paper we study two-dimensional Ising spin glasses, i.e., Ising spin--$1/2$  
magnetic moments interacting via nearest-neighbor bonds $J_{ij}$ of the same strength and 
random sign, drawn from a quenched probability distribution: 
\begin{equation}
P(J_{ij})= p\,\delta (J_{ij}-J_0)+ (1-p)\,\delta (J_{ij}+J_0)\ .
\label{eq:1}
\end{equation}
Our interest focuses on the region of high (low) concentration $p$ ($1-p$) of ferro- 
(antiferro-)magnetic  bonds, where even in  two dimensions one can have order at $T \neq 
0$. A critical line on 
the $T - p$ plane separates paramagnetic and ferromagnetic phases. Furthermore, for
general space dimensionality $d \geq 2$ there  is a second line of 
interest  on  the $T - p$ plane, along which several exact results have been derived, 
known as the {\it Nishimori line} (NL)~\cite{nish81,nish01}. The shape of the
NL is known exactly, and given by
\begin{equation}
e^{-2 J_0/T} = \frac{1-p}{p}\qquad\qquad {\rm (NL,}\ p> \frac{1}{2})\ .
\label{eq:2}
\end{equation}
A multicritical point is present, the {\it Nishimori point} (NP). The NP is 
believed~\cite{ldh88} to be located at the  intersection of the  ferro-paramagnetic 
boundary with the NL. Many
subsequent studies have taken this as a starting assumption, so far with consistent 
results, and we shall do so in the present work. 
As the shape of the phase boundary is known only approximately, 
e.g., from  numerical studies, additional considerations are necessary if one intends to 
pinpoint the exact position of the NP. 

On a square lattice, a conjecture has been put forward~\cite{nn02,mnn03},
to the effect that the NP should belong to a subspace of the $T-p$ plane which is 
invariant under certain duality transformations. For $\pm J$ Ising
systems, the invariant subspace is given by~\cite{nn02,mnn03}:
\begin{equation}
p\,\log_2 (1+e^{-2J_0/T})+(1-p)\,\log_2 (1+e^{2J_0/T})= \frac{1}{2}\  .
\label{eq:3}
\end{equation}
Computing the intersection of Eqs.~(\ref{eq:2}) and~(\ref{eq:3}), the exact location of 
the NP is predicted to be at $p=0.889972 \cdots\,$, $T/J_0 =0.956729 \cdots\,$. This 
agrees well with earlier numerical estimates (though, in some cases, it
is slightly outside estimated error bars). For detailed comparisons see, e.g., 
Ref.~\onlinecite{dqrbs03}.

Very recently~\cite{tsn05}, reasoning along the lines of 
Refs.~\onlinecite{nn02,mnn03} produced a conjectured duality relationship between 
locations of the NP on triangular and honeycomb lattices. By incorporating the
NL condition, Eq.~(\ref{eq:2}), considering lattices 1 and 2 dual of each other,
invoking the replica method with $n$ replicas and taking the quenched limit $ n \to 0$, 
and defining
\begin{equation}
H(p) \equiv -p\,\log_2 p - (1-p)\,\log_2 (1-p)\ , 
\label{eq:4}
\end{equation}
it is predicted that, for mutually-dual systems with quenched randomness,
\begin{equation}
H(p_{1c}) + H(p_{2c}) =1\ .
\label{eq:5}
\end{equation}
Using Monte Carlo simulations, the authors of Ref.~\onlinecite{tsn05} established
that $p_c = 0.930(5)$ for the honeycomb, and $0.835(5)$ for the triangular
lattice. Using Eq.~(\ref{eq:4}), these values imply that $ 0.981 < H(p_{1c}) + H(p_{2c}) 
< 1.042$, consistent with the conjecture Eq.~(\ref{eq:5})~.

Our goal here is twofold: first, to provide accurate checks of the location of the NP
for both lattices, which will allow a more stringent test of Eq.~(\ref{eq:5}); and
second, by invoking universality concepts, to gain more information on the
universality class of the NP, through investigation of suitable critical properties
on both lattices. 

Indeed, although many studies have dealt with the NP on square lattices,
knowledge of the associated scaling indices is still restricted to (sometimes
contradictory) numerical estimates. This is in contrast with the situation for pure
discrete-symmetry systems in two dimensions, where it
has been established that (i) all critical exponents are rational numbers belonging 
to a grid allowed by conformal invariance~\cite{cardy87}, 
and (ii) for each universality class the 
corresponding values have been unambiguously determined from the subset allowed by such 
grid, via additional exact results and/or numerical work. Even when 
(unfrustrated) disorder is introduced, significant progress can be achieved 
(for a recent review see, e.g., Ref.~\onlinecite{bc04} and references therein):
though the connection to rational values of the exponents is lost, estimates
obtained by various (analytical or simulational) methods are usually fairly
consistent.
 
Here we apply numerical transfer-matrix (TM) methods to the spin--$1/2$ Ising
spin glass, on strips of triangular (T) and honeycomb (HC) lattices of widths 
$4 \leq N \leq 13$ sites (T) and $4 \leq N \leq 16$ sites (even values only, HC).
In Sec.~\ref{sec:2}, domain-wall energies are computed, and their finite-size
scaling allows us to estimate both the location $p_c$ of the NP along the NL,
and the correlation-length index, $y_t \equiv 1/\nu$ which governs the spread
of ferromagnetic correlations upon crossing the ferro-paramagnetic phase boundary.
In Sec.~\ref{sec:3}, uniform susceptibilites are calculated, and the associated
exponent ratio $\gamma/\nu$ is evaluated. In Sec.~\ref{sec:4}, we turn to probability
distributions of spin-spin correlation functions, and their moments of assorted orders.
These are shown to obey rather strict conformal-invariance requirements, once
suitable adaptations are made to account for geometric aspects of the TM description  of
T and HC lattices.
Finally, in Sec.~\ref{conc}, concluding remarks are made.

\section{Domain-wall scaling}
\label{sec:2}
For pure two-dimensional systems, the duality between correlation length $\xi$ and 
interface tension $\sigma$ is well-established~\cite{watson}. For an infinite 
strip of width $L$, conformal invariance gives at criticality~\cite{car84}:
\begin{equation}
L\,\sigma_L = \pi\eta\ ,
\label{eq:sigeta}
\end{equation}
where $\eta$ is the decay-of-correlations exponent, and $\sigma_L$ is the domain-wall 
free energy, i.e., the free energy per unit length, in units of $T$,
of a seam along the full length of the strip: for Ising systems, $\sigma_L = 
f^A_L-f^P_L$, with $f^P_L$ ($f^A_L$) being the corresponding free energy  for a strip
with periodic (antiperiodic) boundary conditions across. 
Finite-size scaling properties of $\sigma_L$ have been used in the study of
critical properties of disordered systems as well~\cite{mm84}, including an 
investigation of the NP on a square lattice~\cite{hpp01}.  
With the above definition one has, for non-homogeneous couplings
as is the case here, $\sigma_L = -\ln
(\Lambda_0^A /  
\Lambda_0^P)$ where $\ln \Lambda_0^P$, $\ln \Lambda_0^A$ are the
largest Lyapunov exponents of the TM, respectively with periodic and
antiperiodic boundary conditions across.
  
We have calculated $\Lambda_0^P$, $\Lambda_0^A$ for strips of both T and HC lattices,
usually of length $M=2 \times 10^6$ columns, and widths $N$ as listed in 
Sec.~\ref{intro} (with the exception of $N=16$ for HC). It must be recalled that both 
$L$ in Eq.~(\ref{eq:sigeta}) and the 
correlation length $\xi$ (of which the surface tension is the dual) are actual physical  
distances, in lattice parameter units~\cite{pf84,bww90,bn93,dq00}. Denoting by 
$\Delta$ the column-to-column distance by which the TM progresses in one iteration, 
the usual representations of the T lattice as a square lattice with a single diagonal 
bond, and of the HC as a ``brick" lattice (i.e., with vertical bonds alternately 
missing), imply that
$L=\zeta N$, with $\zeta_T=1$, $\zeta_{HC}=3/2$; $\Delta_T=\sqrt{3}/2$;  $\Delta_{HC}=
\sqrt{3}$ (this latter is because two iterations of the TM are necessary in order to
restore periodicity). The universal quantity $\eta$ is then given by 
\begin{equation}
\eta= \frac{L \sigma_L}{\pi}=\begin{cases}{ 
(2N/\sqrt{3}\pi)\left(\ln\Lambda_0^P 
-\ln \Lambda_0^A\right)\quad {\rm (T)}}\cr
{(N\sqrt{3}/2\pi)\left(\ln\Lambda_0^P
-\ln \Lambda_0^A\right)\quad {\rm (HC)}\ .} 
\end{cases}
\label{eq:etalatt}
\end{equation}
For both lattices we scanned the NL, taking the respective intervals quoted in  
Ref.~\onlinecite{tsn05} as a starting guess for the location of the NP.

\begin{figure}
{\centering \resizebox*{3.4in}{!}
{\includegraphics*{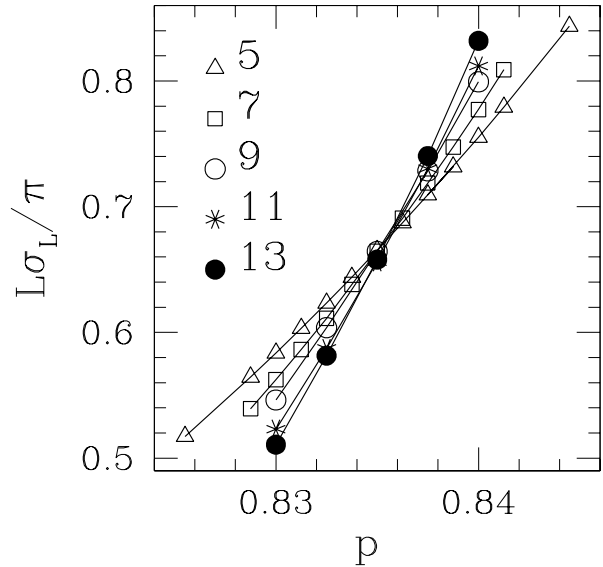}} \par}
\caption{Triangular lattice: domain-wall free energies, 
Eq.~(\protect{\ref{eq:sigeta}}), along the NL (parametrized by $p$, see 
Eq.~(\protect{\ref{eq:2}})). Only data for odd $N$ are shown, in order to avoid
cluttering. Error bars are of order of, or smaller, than symbol sizes.
}
\label{fig:dwt}
\end{figure}

For the T lattice, data for the scaled domain-wall energy are shown  in 
Fig.~\ref{fig:dwt}.  
\begin{figure}
{\centering \resizebox*{3.4in}{!}
{\includegraphics*{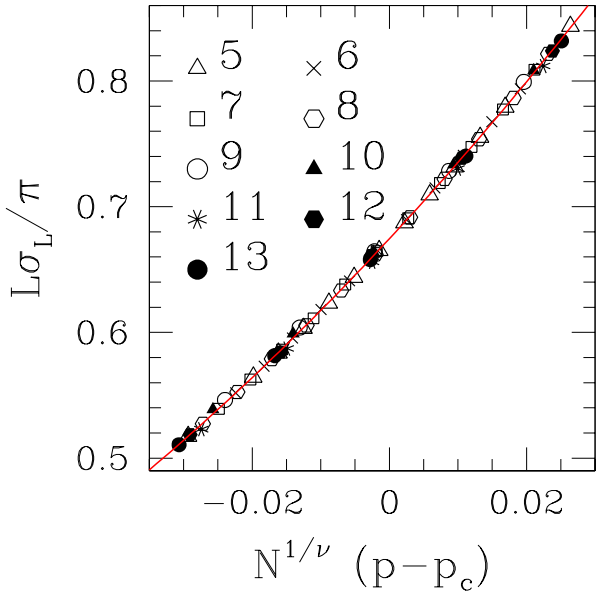}} \par}
\caption{(Color online) Triangular lattice: scaling plot of domain-wall free energies, 
Eq.~(\protect{\ref{eq:sigeta}}), against the finite-size scaling variable,
$N^{1/\nu}\,(p-p_c)$. The central estimates $1/\nu=0.67$, $p_c=0.8355$, have been used.
Full line is quadratic fit to data, from which $\eta=0.674(11)$ (see text).
}
\label{fig:dwtsc}
\end{figure}
Standard finite-size scaling~\cite{barber} suggests that the curves of Fig.~\ref{fig:dwt}
would coincide when plotted against $x \equiv N^{1/\nu}\,(p-p_c)$.  

A quantitative measure of how good the data collapse is can be provided as
follows. For trial values of $(\nu, p_c)$ one calculates the
$\chi^2$ per degree of freedom ($\chi^2_{\rm \ d.o.f.}$) of a fit of the
data to a phenomenological baseline curve $f(x)$ (in the present case,
since the curvature of data was monotonic, we found a parabolic form to be 
satisfactory). As the fractional uncertainties of data points were all of
the same order, we used {\em unweighted} fits, i.e., the $\chi^2_{\rm \
d.o.f.}$ was calculated via
\begin{equation}
\chi^2_{\rm\ d.o.f.} =(N_d-M)^{-1}\sum_{i=1}^{N_d}\left(y_i
-f(x_i)\right)^2\ ,
\label{eq:chi2def}
\end{equation}
where $N_d$ stands for the number of data, $N_d-M$ is the number of degrees
of freedom ($M$ is the number of free parameters), $(x_i,y_i)$
are the data points, and $f(x_i)$ are the values of the fitting
function at the respective $x_i$. The use of uweighted fits is justified
because all data used in each fit have similar fractional uncertainties.
Therefore, the comparative analysis of different fitting parameters
for a specified set of data will not suffer from distortions.
This was the procedure used in all
data collapse analyses in the present work. 

For domain-wall energies on the T lattice, we have found that the best
collapse occurs for $1/\nu =0.67(1)$,
$p_c=0.8355(5)$. For the central 
estimates the $\chi^2_{\rm \ d.o.f.}$ is $3 \times 10^{-6}$. Within the
intervals of confidence given , the $\chi^2_{\rm \
d.o.f.}$ remains below $10^{-5}$. Fig.~\ref{fig:dwtsc} illustrates the quality of
plot obtained, when the central estimates just quoted are used.  A parabolic fit
to the scaled data gives  $\eta=0.674(11)$, where uncertainties in $1/\nu$ and
$p_c$ have been taken into account, in addition to those intrinsic to the fitting
process for fixed values of these parameters.

\begin{figure}
{\centering \resizebox*{3.4in}{!}
{\includegraphics*{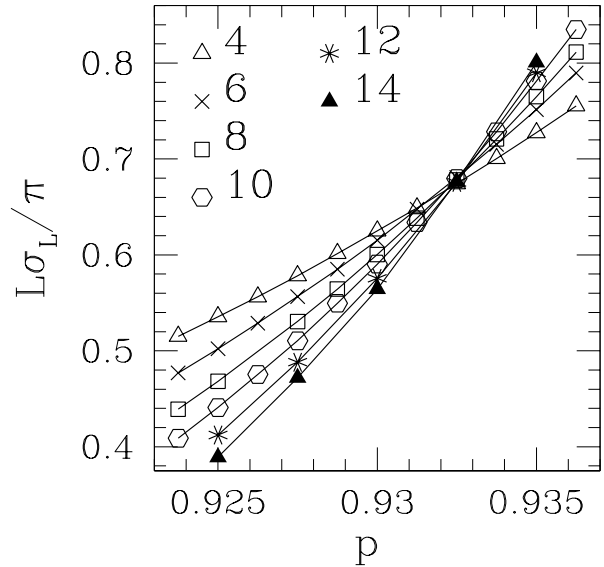}} \par}
\caption{Honeycomb lattice: domain-wall free energies, 
Eq.~(\protect{\ref{eq:sigeta}}), along the NL (parametrized by $p$, see 
Eq.~(\protect{\ref{eq:2}})). Error bars are of order of, or smaller, 
than symbol sizes.
}
\label{fig:dwhc}
\end{figure}

\begin{figure}
{\centering \resizebox*{3.4in}{!}
{\includegraphics*{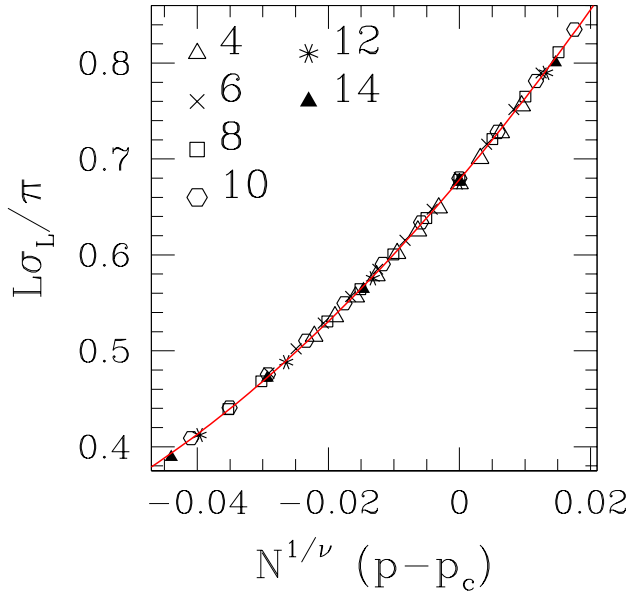}} \par}
\caption{(Color online) Honeycomb lattice: scaling plot of domain-wall
free energies, 
Eq.~(\protect{\ref{eq:sigeta}}), against the finite-size scaling variable,
$N^{1/\nu}\,(p-p_c)$. The central estimates $1/\nu=0.67$, $p_c=0.9325$, 
have been used.
Full line is quadratic fit to data, from which $\eta=0.678(15)$ (see 
text).
}
\label{fig:dwhcsc}
\end{figure}

A similar line of analysis was followed for the HC lattice. 
Fig.~\ref{fig:dwhc} shows the unscaled domain-wall energy data, while
Fig.~\ref{fig:dwhcsc} is a scaling plot for the same data. 
The best collapse occurs for $1/\nu =0.67(1)$, $p_c=0.9325(5)$. For the central 
estimates the $\chi^2_{\rm \ d.o.f.}$ is $7 \times 10^{-6}$. Within the intervals
of confidence given, the $\chi^2_{\rm \ d.o.f.}$
remains below $2 \times 10^{-5}$. An estimate of $\eta$ from parabolic
fits, with the same considerations used for the T lattice, gives $\eta=0.678(15)$. 

The above estimates of $p_c$ for T and HC lattices, when plugged into Eq.~(\ref{eq:4}), 
result in:
\begin{equation}
H(p_{1c}) + H(p_{2c}) =1.002(3)\ .
\label{eq:hest}
\end{equation} 
This improves on the accuracy of the estimate given in Ref.~\onlinecite{tsn05}
by one order of magnitude, while still being compatible with the
prediction Eq.~(\ref{eq:5}). We view this agreement as a strong indication of
plausibility of the conjecture exhibited in  Ref.~\onlinecite{tsn05}.

As regards the correlation-length exponent, our estimate $\nu=1.49(2)$
is incompatible with $\nu=1.33(3)$ quoted from the same sort of domain-wall
scaling analysis applied to the NP on a square lattice~\cite{hpp01}, but agrees
well with $\nu=1.50(3)$, found from mapping into a network model for disordered
noninteracting fermions, via TM~\cite{mc02a}.

Turning now to the exponent $\eta$ given in Eq.~(\ref{eq:sigeta}), it has been
recalled, e.g., in Ref.~\onlinecite{mc02b}, that in the presence of
disorder,
the scaling indices of the disorder correlator (i.e., the interfacial tension)
differ from those of its dual, the order correlator (namely, spin-spin
correlations). Nevertheless, the constraints of conformal
invariance still hold, with the result that the amplitude of the domain wall
energy remains a {\em bona fide} universal quantity~\cite{mc02b}. For a square
lattice, recent estimates give $\eta=0.691(2)$~\cite{hpp01,mc02a,mc02b}.
This is slightly outside the error bars quoted here for the T lattice, but within
the uncertainty given for HC data.

\section{Uniform susceptibilities}
\label{sec:3}
We calculated uniform zero-field susceptibilities along the NL for both T and HC
lattices, similarly to previous investigations on the square lattice~\cite{sbl3}. 
For the finite differences used in numerical differentiation, we used a field step
$\delta h=10^{-4}$ in units of $J_0$. As in Sec.~\ref{sec:2}, we took the 
respective intervals quoted in Ref.~\onlinecite{tsn05} as a starting guess for the
location of the NP.

Finite-size scaling arguments~\cite{barber} suggest a form
\begin{equation}
\chi_N = N^{\gamma/\nu}\,f\left( N^{1/\nu}(p-p_c)\right)\ ,
\label{eq:chisc}
\end{equation}
where $\chi_N$ is the  finite-size susceptibility, and $\gamma$ is the susceptibility
exponent. In order to reduce the number of fitting parameters, we kept $1/\nu$ and $p_c$
fixed at their central estimates obtained in Sec.~\ref{sec:2}, and allowed 
$\gamma/\nu$ to vary. 

\begin{figure}
{\centering \resizebox*{3.4in}{!}
{\includegraphics*{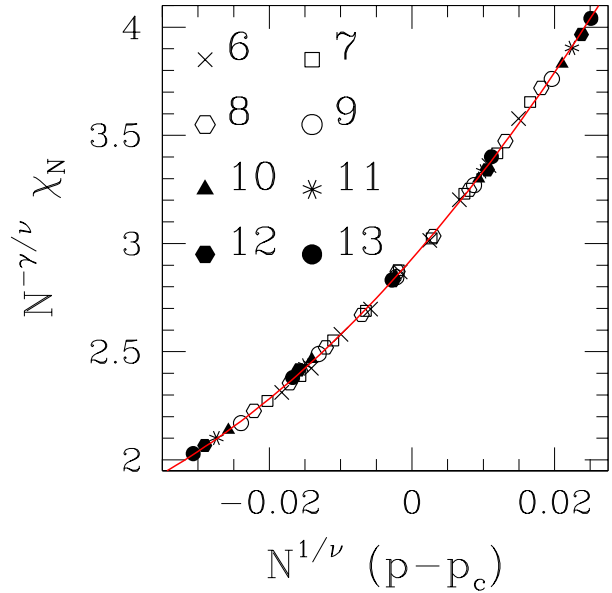}} \par}
\caption{(Color online) Triangular lattice: scaling plot of uniform zero-field
susceptibilities, Eq.~(\protect{\ref{eq:chisc}}). The central estimates
$1/\nu=0.67$, $p_c=0.8355$, $\gamma/\nu=1.795$, have been used.
Full line is quadratic fit to data.
}
\label{fig:chitsc}
\end{figure}

Within this framework, our best fit for the T lattice was for
$\gamma/\nu=1.795(20)$, as shown in Fig.~\ref{fig:chitsc}. 
For the central 
estimate the $\chi^2_{\rm \ d.o.f.}$ is $1 \times 10^{-4}$. Within the intervals
of confidence given, the $\chi^2_{\rm \ d.o.f.}$
remains below $3 \times 10^{-4}$. These deviations are one and a half
orders of magnitude larger than the corresponding ones for domain-wall scaling
(see Sec.~\ref{sec:2}). 

\begin{figure}
{\centering \resizebox*{3.4in}{!}
{\includegraphics*{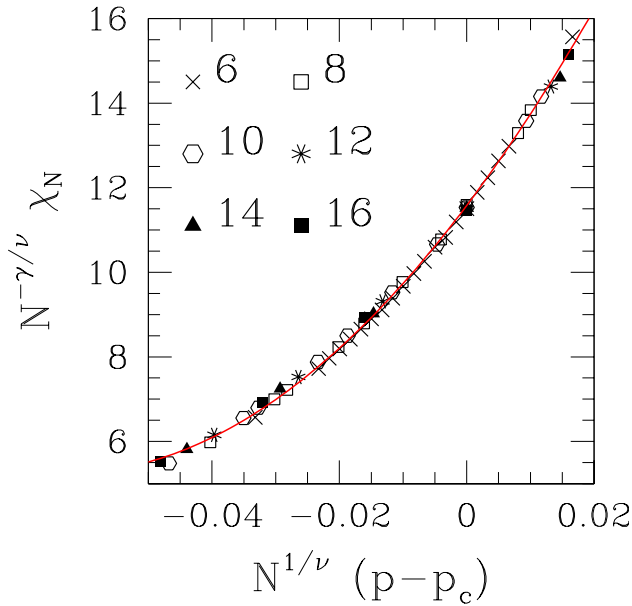}} \par}
\caption{(Color online) Honeycomb lattice: scaling plot of uniform zero-field
susceptibilities, Eq.~(\protect{\ref{eq:chisc}}). The central estimates
$1/\nu=0.67$, $p_c=0.9325$, $\gamma/\nu=1.80$, have been used.
Full line is quadratic fit to data.
}
\label{fig:chihcsc}
\end{figure}

We repeated the same steps for the HC lattice, with the results displayed in
Fig.~\ref{fig:chihcsc}. The best fit now was for $\gamma/\nu=1.80(4)$. For the
central  estimate the $\chi^2_{\rm \ d.o.f.}$ is $1 \times 10^{-2}$, two orders of
magnitude larger than for the T lattice. The lower quality of
adjustment can be witnessed visually. Within the intervals
of confidence given, the $\chi^2_{\rm \ d.o.f.}$ remains below $3 \times 10^{-2}$. 
Though the central estimates for the T and HC lattices are very close, the corresponding
error bars differ by a factor of two. For the square lattice, we quote
$\gamma/\nu=1.80(2)$~\cite{sbl3}, compatible with both values found here.

\section{Correlation functions}
\label{sec:4}
Our study of correlation functions is based on previous work for the square
lattice~\cite{dqrbs03}. We recall the following property, which has been
shown to hold on the NL, for correlation functions $ C_{ij}$ between Ising
spins $\sigma_i$, $\sigma_j$~\cite{nish81,nish01,nish86,nish02}:
\begin{equation}
[\,C_{ij}^{\,(2\ell +1)}] \equiv [\,\langle \sigma_i \sigma_j\rangle^{2\ell 
+1}] = [\,C_{ij}^{\,(2\ell +2)}] \equiv
[\,\langle \sigma_i \sigma_j \rangle^{2\ell +2} ]\ ,
\label{eq:momsc}
\end{equation}
where angled brackets indicate the usual thermal average,  square brackets
stand for configurational averages over disorder, and $\ell = 0,1, 2, \dots$.
Denoting by $P(C_{ij})$ the probability distribution function for the $C_{ij}$, the 
pairing of  successive odd and even moments predicted in  Eq.~(\ref{eq:momsc}) 
implies  that  $P^\prime(C_{ij}) \equiv (1-C_{ij})\,P(C_{ij})$ must be an even 
function of $C_{ij}$, everywhere on the NL~\cite{dqrbs03}. We have explicitly
checked that this constraint is obeyed by the distributions generated for the
T and HC lattices, within the same degree of accuracy as reported in 
Ref.~\onlinecite{dqrbs03} for the square lattice.  We shall not deal 
directly with the $P(C_{ij})$ in what follows; instead, we concentrate 
on the scaling of their assorted moments $m_{\,k} \equiv [\,C_{ij}^{\,k}]$ , 
especially in connection with their conformal-invariance properties. 

In contrast to the symmetry exhibited in 
Eq.~(\ref{eq:momsc}) which holds everywhere on the NL, conformal invariance is
expected only where the NL crosses the phase boundary, i.e., at the NP.

For pure Ising systems on a strip of width $L$ of a square lattice,
with periodic boundary conditions
across, conformal invariance implies that at criticality, the correlation 
function between spins located respectively at the origin and at $(x,y)$ 
behaves as~\cite{cardy87}:
\begin{equation}
C_{xy}^{\rm pure}\sim \left[ \frac{\pi/L}{\left[ \sinh^2 (\pi
x/L)+ \sin^2 (\pi y/L)\right]^{1/2}} \right]^{\eta} \  ,\  \eta =1/4\ .
\label{eq:conf-inv}
\end{equation}
For the T and HC lattices, the same is true, provided that the actual, i.e.,
geometric site coordinates along the strip are used in Eq.~(\ref{eq:conf-inv}).
Thus, from the representation of the T lattice as a square (SQ) lattice with a single 
diagonal bond, and of the HC as a ``brick" lattice, the respective SQ-like integer 
coordinates $(i,j)$ transform respectively into
\begin{eqnarray}
x = \frac{\sqrt{3}}{2}\,i\ ;\ y=j -\frac{1}{2}\,i\qquad {\rm (T)}\ ;
\nonumber \\
x=\sqrt{3}\,i\ ;\ y=j +\bigl[\,\frac{j+1}{2}\,\bigr]\qquad {\rm (HC)}\ ,
\label{eq:geoc}
\end{eqnarray}
where $[X]$ denotes the largest integer contained in $X$. Recall that $L=N$ (T);
$L=3N/2$ (HC), as explained above. 
With $R \equiv (x^2+y^2)^{1/2}$, the proportionality factor in 
Eq.~(\ref{eq:conf-inv}) can be obtained from exact 
results ($L,R \to \infty,\ R \ll L$), $C_{R} = A_{\rm X}/R^{1/4}$, where $A_{\rm 
X}=0.703\,38 \dots$, $0.668\,65 \dots$, 
$0.767\,04 \dots$ respectively for  X = SQ, T, HC~\cite{hs68}. 
\begin{figure}
{\centering \resizebox*{3.4in}{!}
{\includegraphics*{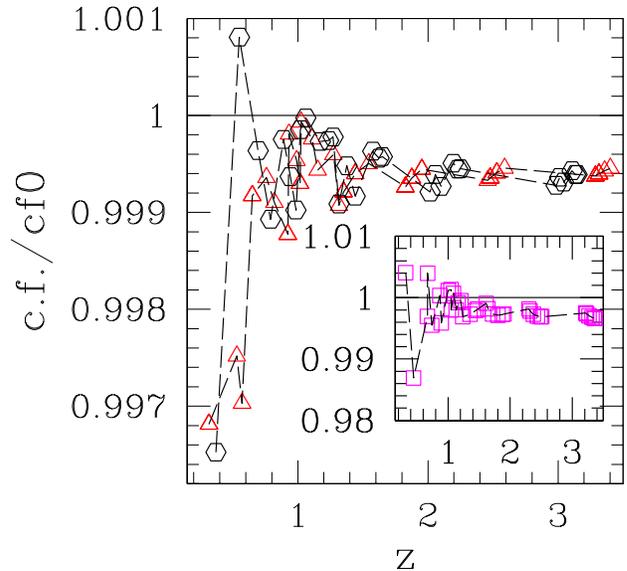}} \par}
\caption{(Color online) Pure systems: ratio of calculated correlation functions at the
critical point (c.f.) to asymptotic form given in Eq.~(\protect{\ref{eq:conf-inv}}) 
(cf0), against $z \equiv (\sinh^2 (\pi x/L)+ \sin^2 (\pi y/L))^{1/2}$. Main diagram:
T lattice (triangles), HC lattice (hexagons). Insert: SQ lattice. Strip 
width $N=10$ sites, for all cases. 
}
\label{fig:cfpure}
\end{figure}
Though strictly 
speaking Eq.~(\ref{eq:conf-inv}) is an asymptotic form, for the SQ lattice discrepancies 
are already very small at short distances~\cite{dqrbs03}, and are even smaller for T and 
HC, as illustrated in Fig.~\ref{fig:cfpure}. The horizontal axis in the Figure is the
argument $z \equiv (\sinh^2 (\pi x/L)+ \sin^2 (\pi y/L))^{1/2}$ of
Eq.~(\ref{eq:conf-inv}). The range of $z$ depicted corresponds to $x/L 
\lesssim 0.6$, i.e.  (for strip width $N=10$ sites) up to, respectively, 5(HC), 6(SQ), or 
7 (T) full iterations of the TM. For larger $x/L$ the angular dependence of $z$ (through 
$y$) becomes less than one part in $10^2$. For $z \lesssim 1$ the discrepancy from 
Eq.~(\ref{eq:conf-inv}) is at most $0.4\%$ for both T and HC, while in the worst case
for SQ, namely $(x,y)$=(1,1), it reaches $1.3\%$. For $1 < z < 3$, the difference is
$ < 0.1\%$ for T and HC, and  $< 0.3\%$ for SQ.   

The above analysis of conformal invariance of pure-system correlation functions
indicates that, should similar trends hold at the NP of spin glasses, estimates of
associated critical indices for T and HC lattices would behave more smoothly than for 
SQ (since they rely on fits of numerically-calculated correlations
to Eq.~(\ref{eq:conf-inv}), with $\eta$ as an adjustable parameter). As in earlier
work~\cite{dqrbs03}, we concentrate on short-distance correlations, i.e., where the 
argument $z$ is strongly influenced by $y$. Such a setup is especially convenient in 
order to probe the angular dependence predicted in Eq.~(\ref{eq:conf-inv}), which
constitutes a rather stringent test of conformal invariance properties.

We now turn to the quantitative analysis of the behavior of assorted moments $m_i$ of the 
correlation-function distribution, against $z$. Bearing in mind Eq.~(\ref{eq:momsc}), 
and following Ref.~\onlinecite{dqrbs03}, our 
goal is to extract the  decay-of-correlations exponents $\eta_{\,2j+1}$, via fits of our
data to the form $m_{\,2j+1} \sim z^{-\eta_{\,2j+1}}$. 

When one attempts such fits,
several likely sources of  uncertainty  are  present, on which we now comment. First, 
one has the 
finite width $N$ of the strips used. In Ref.~\onlinecite{dqrbs03}, an extensive analysis 
of this point was undertaken, with the conclusion that, e.g. for $N=10$, finite-width 
effects are
already essentially subsumed in the explicit $L$ (i.e., $N$) dependence of 
Eq.~(\ref{eq:conf-inv}), thus higher-order finite-size corrections most likely do not
play a significant role. We shall assume that this is the case here as well, and
restrict ourselves to $N=10$ for both T and HC lattices.
Second, the finite length $M$ of strips implies that averaged values will fluctuate from
sample to sample. Though the distribution itself (of, e.g., correlation 
functions) displays an intrinsic width which is a non-vanishing feature connected
to the lack of self-averaging present at criticality,
the average moments of the distribution behave in the expected manner, namely,
their sample-to-sample fluctuations approach zero roughly as $\sqrt{M}$ with
increasing sample length $M$~\cite{dqrbs96,ts94}. Therefore, from a set of runs at 
assorted small values of $M$, one can infer what effect sample-to-sample fluctuations 
will have on results for larger $M$. In the calculation of results shown below, we have 
used $M=10^7$, which implies a total of $M^\prime=3.3 \times 10^6$ 
non-overlapping samples for our correlation-function statistics (because each sample
needs three full iterations of the TM, in order to scan the set of lattice points
of interest). For such value of $M$, the  estimation procedure just outlined predicts 
fluctuations of order $0.1\%$, at most.
 
Finally, one has the uncertainty in the location of the critical point.
We  have found that, in the  present case, this is the main source of  
uncertainties for our data. Thus, e.g., with $p_c=0.8355(5)$ for the T lattice, averaged 
moments $m_{\,2j+1}$ taken at at the central estimate differ from those calculated at the 
edge of the error bar, by an amount increasing systematically with $j$, from $\sim 0.7\%$ 
for $j=0$, to $\sim  1.5\%$ for $j=3$. For HC, deviations follow the same trend 
against $j$ but are slightly larger, ranging from  $\lesssim 1\%$
for $j=0$, to $\lesssim  2\%$ for $j=3$.

\begin{figure}
{\centering \resizebox*{3.4in}{!}
{\includegraphics*{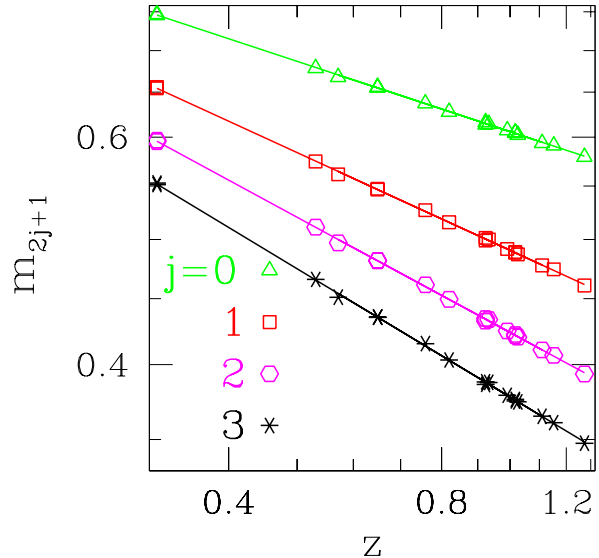}} \par}
\caption{(Color online) Triangular lattice:
double-logarithmic plot of odd moments of the correlation-function
distribution $P(C_{xy})$ against $z \equiv ( \sinh^2 (\pi x/L)+ \sin^2
(\pi y/L))^{1/2}$ (see Eqs.~(\protect{\ref{eq:conf-inv}}) and 
~(\protect{\ref{eq:geoc}})). Straight lines are unweighted least-squares fits to data.
Data taken at $p=0.8355$ for strip width $N=10$, $M^\prime=3.3 \times 10^6$
non-overlapping samples in all cases.
}
\label{fig:etat}
\end{figure}

In Fig.~\ref{fig:etat} we show data for the T lattice, taken at our central estimate for 
the location of the NP, $p=0.8355$. The error bars, associated mainly to the uncertainty
in $p_c$, as just discussed, are at most of order of the symbol sizes.
\begin{figure}
{\centering \resizebox*{3.4in}{!}
{\includegraphics*{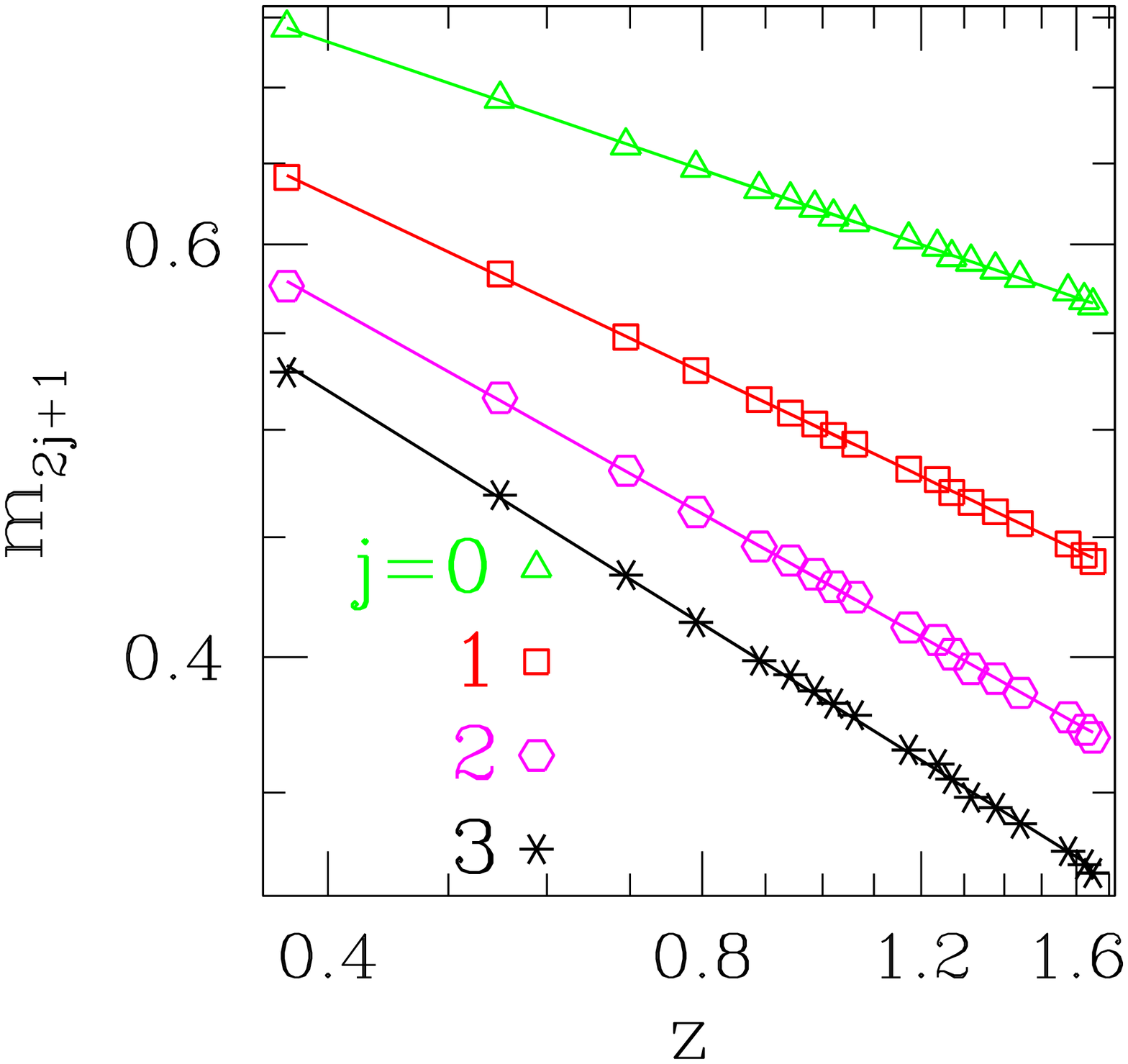}} \par}
\caption{(Color online) Honeycomb lattice:
double-logarithmic plot of odd moments of the correlation-function
distribution $P(C_{xy})$ against $z \equiv ( \sinh^2 (\pi x/L)+ \sin^2
(\pi y/L))^{1/2}$ (see Eqs.~(\protect{\ref{eq:conf-inv}}) and 
~(\protect{\ref{eq:geoc}})). Straight lines are unweighted least-squares fits to data.
Data taken at $p=0.9325$ for strip width $N=10$, $M^\prime=3.3 \times 10^6$
non-overlapping samples in all cases.
}
\label{fig:etahc}
\end{figure}
Fig.~\ref{fig:etahc} exhibits data for the HC lattice. Pertinent comments are similar
to those made above for the T lattice.
\begin{table}
\caption{\label{t1}
Estimates of exponents $\eta_{\,2j+1}$, from least-squares fits of
averaged odd moments of
correlation-function distributions. Data for $N=10$ and $z \lesssim 1.6$, assuming  
$m_{2j+1} \sim z^{-\eta_{\,2j+1}}$. T: triangular lattice (this work); HC:
Honeycomb lattice (this work); SQ: square lattice, calculated at the conjectured exact 
location of the NP, see Eq.~(\protect{\ref{eq:3}}) (Ref.~\protect{\onlinecite{dqrbs03}}).
Last two columns: square lattice, authors as quoted.
}
\vskip 0.2cm
\begin{ruledtabular}
\begin{tabular}{@{}llllll}
$j$ & T & HC & SQ & Ref.~\protect{\onlinecite{hpp01}} & 
Ref.~\protect{\onlinecite{mc02a}}\\
0 &  0.181(1) & 0.181(1) & 0.1854(17) & 
0.1854(19)& 0.183(3) \\
1 &  0.251(1) & 0.252(1) & 0.2556(20) & 
0.2561(26)& 0.253(3)\\
2 &  0.297(2) & 0.296(2) & 0.300(2)  & 0.3015(30) & -- \\
3 &  0.330(2) & 0.329(3) & 0.334(3) & 0.3354(34) & -- \\
\end{tabular}
\end{ruledtabular}
\end{table}

In Table~\ref{t1} we give numerical results of the fits illustrated in 
Figs.~\ref{fig:etat} and~\ref{fig:etahc}. Though T and HC estimates are quite consistent 
with each other, and with the results of Ref.~\onlinecite{mc02a}, for $j=0$ and $1$ both 
fall slightly below their SQ counterparts given in Refs.~\onlinecite{dqrbs03,hpp01}.
For $j=2$ and $3$, as a consequence of generally wider error bars, all estimates are
broadly compatible with one another.

\section{Discussion and Conclusions}
\label{conc}
We have used domain-wall scaling techniques in Sec.~\ref{sec:2} to determine the location 
of the Nishimori point of Ising spin glasses on both the T and HC lattices.
Probing the temperature--concentration plane along the Nishimori line, we have obtained 
well-behaved curves of interfacial free energy; with the help of standard finite-size
scaling techniques, we have extracted the estimates $p_c=0.8355(5)$ and $p_c=0.9325(5)$
respectively for the location of the Nishimori point on T and HC lattices.
As a consequence of this, we have been able to refine the estimate of the quantity
$H(p_{1c}) +H(p_{2c})$ (see Eqs.~(\ref{eq:4}) and~(\ref{eq:5})), which has been 
conjectured in Ref.~\onlinecite{tsn05} to be 
exactly unity. Indeed, our result given in Eq.~(\ref{eq:hest}) is $1.002(3)$,
which gives strong support to the conjecture cited.

Furthermore, interfacial free energy data have allowed us to estimate the
correlation-length exponent to be $\nu=1.49(2)$, in very good agreement with 
$\nu=1.50(3)$ from a mapping of the problem into a network model for disordered
noninteracting fermions~\cite{mc02a}, but incompatible with $\nu=1.33(3)$ from
a TM treatment, presumably very similar to the present one, for the SQ
lattice~\cite{hpp01}. 

\begin{figure}
{\centering \resizebox*{3.4in}{!}
{\includegraphics*{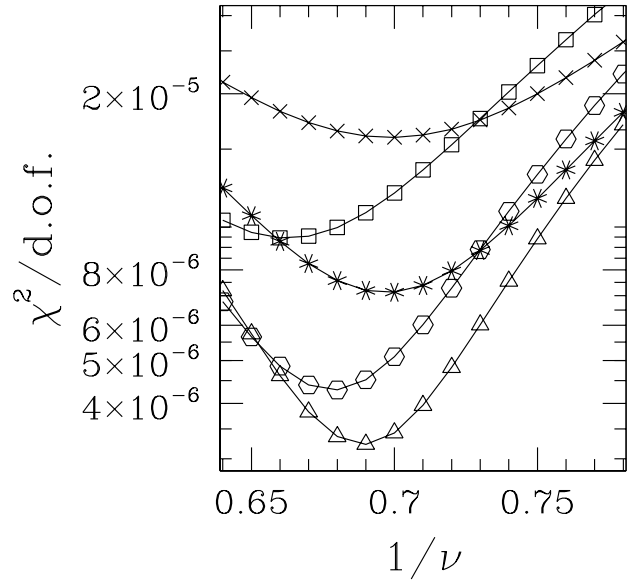}} \par}
\caption{Square lattice:
semi-logarithmic plots of $\chi^2_{\rm\ d.o.f}$ for (unweighted) 
quadratic fits of domain-wall energies on SQ lattice to a dependence on
the finite-size scaling variable $N^{1/\nu}\,(p-p_c)$, against $1/\nu$.
Each set of data
corresponds to fixed $p_c$, as follows: triangles, $p_c=0.889972$
(conjectured exact, Refs.~\protect{\onlinecite{nn02,mnn03}});
squares, $p_c=0.8894$; hexagons, $p_c=0.8897$; stars, $p_c=0.8903$;
crosses, $p_c=0.8906$ (central estimate of
Ref.~\protect{\onlinecite{hpp01}}). Strip widths $4 \leq N \leq 12$.
}
\label{fig:chi2sq}
\end{figure}

In order to investigate whether this latter disagreement might indicate
a lattice-dependent breakdown of universality, we calculated domain-wall 
free energies on the SQ lattice as well. Strip widths $N=4-12$ (both
even and odd) were used, again with $M=2\times 10^6$ columns (except for
$N=12$ where $M=1\times 10^6$). We scanned  the region of the NL
comprising $0.88 \lesssim p \lesssim 0.90$, which includes both the
conjectured exact location of the NP~\cite{nn02,mnn03},
namely $p_c=0.889972 \cdots$, and the estimate given in
Ref.~\onlinecite{hpp01}, $p_c=0.8906(2)$. We found that scaled data
collapse more smoothly for $p_c$ and $\nu$ respectively close to
$0.889972\cdots$ and $1.5$, rather than the values quoted in
Ref.~\onlinecite{hpp01}. This is illustrated in Fig.~\ref{fig:chi2sq},
which exhibits the  $\chi^2_{\rm\ d.o.f}$ for unweighted quadratic fits
of scaled domain-wall energies to a dependence on
the finite-size scaling variable $N^{1/\nu}\,(p-p_c)$, plotted against
$1/\nu$. Each set of data corresponds to fixed $p_c$, see caption to the
Figure. We obtain $\nu=1.45(8)$, $p_c=0.8900(5)$, where the intervals of
confidence given reflect the region in $(\nu, p_c)$ parameter space in
which the $\chi^2_{\rm\ d.o.f}$  
remains below $\sim 1.5$ times its overall minimum.
Though the error bar for $\nu$ is double that for T and HC lattices,
the present estimate still encompasses the respective results for both,
while excluding $\nu=1.33$. For the domain-wall energy amplitude,
we quote $\eta =0.665(10)$, slightly lower than, but still compatible with,
the values found in Sec.~\ref{sec:2}.
We conclude that our domain-wall energy data fully support a picture
of universal (i.e., lattice-independent) behavior at the NP of T, HC, and
SQ lattices. For all three lattices the correlation-length exponent is
consistent with $\nu=1.50(3)$ of Ref.~\onlinecite{mc02a}, but most likely
excludes $\nu=1.33(3)$ of Ref.~\onlinecite{hpp01}.

Our data for the uniform susceptibility, exhibited in Sec.~\ref{sec:3}, 
do not scale as smoothly as the domain-wall energies. Nevertheless, the application
of finite-size scaling ideas yields estimates for the exponent ratio $\gamma/\nu$
which strongly support universal behavior at the NP, for T, HC, and SQ lattices.
We recall that early work characterized the transition at the NP as compatible with
the universality class of random percolation (see, e.g., 
Refs.~\onlinecite{sbl3,hpp01,dqrbs03}
for discussions of this point). In this context, we  
note that even our most accurate single result, namely $\gamma/\nu=1.795(20)$ for
the T lattice, does not rule out the percolation value~\cite{sa94} 
$(\gamma/\nu)_p=43/24=1.7917 \dots$. 
However, as explained below, consideration of the full set of results
obtained here does support a scenario which rules out percolation-like behavior.

Next, we turn to the investigation of correlation functions in Sec.~\ref{sec:4}. The 
rapid convergence of T and HC results towards the asymptotic form, illustrated in
Fig.~\ref{fig:cfpure} for pure systems, has translated to some extent into a discernible 
improvement on the accuracy of estimates for the disordered case. 
As explained above, for spin glasses on T and HC lattices the uncertainty in the location 
of  the NP is the  main source of fluctuations in numerically-calculated quantities. 
Thus, the relatively small uncertainties shown in Table~\ref{t1} show that
the former effect compensates for the noise associated to the latter, at least partially.
Compare, e.g., the T and HC columns with that for data taken at the conjectured exact 
location of the NP on SQ~\cite{dqrbs03}.
 
The overall  picture summarized in Table~\ref{t1} clearly points towards universality of 
the several (multifractal)~\cite{mc02a,mc02b} decay-of-correlation exponents. The small 
discrepancies observed, for $j=0$ and $1$, between the T and HC estimates, and a 
subset of those obtained earlier for SQ, are likely to depend on details of the 
respective fitting procedures. One must note, however, that the $j=0$ and $1$ T and HC 
estimates are  consistent with those derived in Ref.~\onlinecite{mc02a}. This is 
similar to the case for the exponent $\nu$, in which our own result is compatible with 
the value found in Ref.~\onlinecite{mc02a}, and not with that given in 
Ref.~\onlinecite{hpp01}.

Focusing now on $j=0$, an unweighted average of all results of the corresponding line
in Table~\ref{t1} gives $\eta_1= 0.183(3)$. Considering the scaling relation
$\gamma/\nu=2-\eta_1$, one gets $\gamma/\nu=1.817(3)$, which excludes $(\gamma/\nu)_p$
by a broad margin. Therefore, we quote the set of exponents $\nu=1.49(2)$, 
$\gamma=2.71(4)$ $\eta_1= 0.183(3)$, distinctly different from the percolation 
values~\cite{sa94} $\nu_p=4/3$, $\gamma_p=43/18$, $\eta_p=5/24$. 

In summary, we have (i) produced accurate estimates of the location of the NP on 
T and HC lattices, which provide strong evidence in support of the conjecture expressed
in Eq.~(\ref{eq:5}); (ii) confirmed that the critical properties of the NP in 
two-dimensional systems are universal in the expected sense; and (iii) provided further 
evidence that such properties belong to a distinct universality class from that of
percolation.

As a final remark, we note that our discussion has been restricted to critical behavior
upon crossing the ferro-paramagnetic phase boundary. The  critical properties along the
boundary line are of interest as well~\cite{sbl3,mc02a}, and their investigation
on the T and HC lattices would be a natural continuation of the present work. 

\begin{acknowledgments}

This research  was partially supported by 
the Brazilian agencies CNPq (Grant No. 30.0003/2003-0), FAPERJ (Grant
No. E26--152.195/2002), FUJB-UFRJ, and Instituto do Mil\^enio de
Nanoci\^encias--CNPq.
\end{acknowledgments}

\bibliography{biblio}  
\end{document}